\documentclass{emulateapj}
\usepackage{lscape}

\bibpunct{(}{)}{;}{a}{}{,}
\usepackage{psfig}

\newcommand{\chandra}{{\emph{Chandra}}}

\newcommand{\flux}{\hbox{erg~cm$^{-2}$~s$^{-1}$}}

\newcommand{\lumin}{\hbox{erg~s$^{-1}$}}
\newcommand{\OIII}{[O~{\textsc{iii}}]}

\begin{document}

\shortauthors{Peterson et al.}
\shorttitle{Artificial $z$ = 0.3 Chandra Observations} 

\title{Seyfert Galaxies and the Hard X-ray Background: Artificial {\em CHANDRA} Observations of \lowercase{$z$} = 0.3 Active Galaxies}
\author{
K.\ C. Peterson,\altaffilmark{1}
S.\ C. Gallagher,\altaffilmark{1}
A.\ E. Hornschemeier,\altaffilmark{2}
M.\ P. Muno,\altaffilmark{1}
and E.\ C. Bullard\altaffilmark{3}
}
\altaffiltext{1}{Department of Physics \& Astronomy, University of
  California -- Los Angeles, Mail Code 154705, 475 Portola Plaza, Los
  Angeles CA, 90095--4705; kptrson@astro.ucla.edu, sgall@astro.ucla.edu, 
  mmuno@astro.ucla.edu}

\altaffiltext{2}{Laboratory for X-ray Astrophysics, NASA Goddard Space Flight Center,
Code 662.0, Greenbelt, MD 20771; annh@milkyway.gsfc.nasa.gov}
\altaffiltext{3}{Center for Space Research, Massachusetts Institute of
Technology,  77 Massachusetts Avenue, Cambridge, Massachusetts 02139; ebullard@mit.edu}

\begin{abstract}
Deep X-ray surveys have resolved much of the X-ray background radiation below 2~keV into discrete sources, but 
the background above 8~keV remains largely unresolved. 
The obscured (type 2) Active Galactic Nuclei (AGNs) that are expected to dominate 
the hard X-ray background have not yet been detected in sufficient numbers 
to account for the observed background flux.
However, deep X-ray surveys have revealed large numbers of faint quiescent and starburst galaxies 
at moderate redshifts. In hopes of recovering the missing AGN population, it has been suggested that 
the defining optical spectral features of low-luminosity Seyfert nuclei at large distances
may be overwhelmed by their host galaxies, causing them to appear optically quiescent in deep surveys.
We test this possibility by artificially redshifting a sample of 23 nearby, well-studied active 
galaxies to $z = 0.3$, testing them for X-ray AGN signatures and comparing them
to the objects detected in deep X-ray surveys. We find that these redshifted galaxies
have properties consistent with the deep field ``normal'' and ``optically bright, X-ray faint'' (OBXF) 
galaxy populations, supporting the 
hypothesis that the numbers of AGNs in deep X-ray surveys are being underestimated, and suggesting 
that OBXFs should not be ruled out as candidate AGN hosts that could contribute to the hard X-ray background source population.

\end{abstract}

\keywords{galaxies: active --- galaxies: Seyfert --- X-rays: diffuse background --- X-rays: galaxies}

\section{INTRODUCTION}
One of the fundamental pursuits of deep X-ray surveys is the identification of the
sources making up the observed X-ray background radiation \citep{araa}.
Because the X-ray powerhouses of the universe are energetically accreting super-massive
black holes, or Active Galactic Nuclei (AGNs),
this issue is inherently linked to another important topic: an 
understanding of the cosmic history of super-massive black hole (SMBH) accretion activity.
X-ray surveys are arguably among the best methods for obtaining 
the least-biased samples of AGNs because X-rays are able to penetrate large column densities of gas and dust,
and because dilution by host galaxy light at 
X-ray energies is believed to be minimal.

Using sensitive instruments such as the \emph{Chandra X-ray Observatory} \citep{chandra_ref}, 
astronomers have resolved $\approx$$90\%$ of the 0.5--2~keV background radiation 
into discrete sources. The dominant contributors to the background in this soft energy band have been 
identified via optical spectroscopy as 
type 1 Seyfert galaxies and quasars, i.~e., moderate-luminosity ($L_{0.5-10~\rm keV} \sim 10^{43}$--$10^{45}$ \lumin) AGNs 
recognized by their broad permitted emission lines in the optical \citep{rosat,lehmann,wors05}. 
Deep surveys reveal that the integral number counts of resolved sources at soft energies flatten at fluxes
below $\approx$$10^{-14}$ \flux, confirming 
that most of the population has already been detected \citep{rosati}.
Above 8~keV, however, only $\approx$$50\%$ of the 
X-ray background has been resolved, and, in the 5--10 keV energy band, no turn-over in number counts is observed out to 
the flux limits of current X-ray surveys \citep{wors05,araa}.
This suggests that a substantial fraction of the hard X-ray background is coming from sources 
that are unrepresented at soft energies; indeed, 14--20\% of the 2--8~keV sources are undetected in the 
0.5--2~keV band in deep X-ray surveys \citep{alex03,lehmer05}.
Furthermore, while the integrated light from the 1--10~keV background can be fit with a power-law model 
of the form $F(E) \propto E^{-\Gamma}$, with photon index $\Gamma = 1.4$ \citep{deluca}, the combined spectrum of the 
resolved X-ray background components is much softer. 
\citet{wors04} measured an average 0.2--12~keV photon index of $\Gamma \sim 1.75$ for the 
resolved X-ray sources, with the slope being steeper ($\Gamma \sim 1.80$) at hard energies and shallower 
($\Gamma \sim 1.65$) at soft energies.
They also found that the photon index is harder (with a flatter spectral slope) for the X-ray faint resolved sources than for the 
X-ray bright sources. All of this points to a population of as-yet undetected, faint, hard objects contributing 
significantly to the hard ($E > 8$~keV) X-ray background.

Type 2 (narrow optical emission line) Seyfert galaxies have emerged as the preferred candidates 
for this unresolved population.
While the intrinsic luminosities of Seyfert 2 nuclei can be comparable to Seyfert 1 nuclei, 
their continuum emission from the optical to the soft X-rays is highly obscured. 
Many of these objects are optically thick to Compton scattering 
(``Compton thick'', $N_H$~$>$~1.5~$\times$~10$^{24}$~cm$^{-2}$), 
and are therefore so heavily absorbed at soft X-ray energies that only reprocessed, indirect X-rays are observable, and the 
resulting spectra are hard, consistent 
with the expected properties of the unresolved background population.
Even objects with smaller column densities may be highly obscured; for example, the observed 2--10~keV flux of a $\Gamma = 2$ object 
with $N_H$ = 5 $\times$ 10$^{23}$~cm$^{-2}$ would be only one-third of its unabsorbed flux.
Above $\sim$10~keV, type 2 AGNs may appear much brighter than at soft energies because their
higher energy emission is much less affected by obscuring gas 
\citep[e.g.][]{beckmann}. 
Therefore, their contribution to the X-ray background above 10~keV cannot be estimated with a simple 
extrapolation of their lower energy spectra. It is 
important to recognize the obscured AGN character of these objects so that their emerging hard X-ray flux can be 
included in accounting for the X-ray background.

As it turns out, X-ray observers have not been able to identify a sufficiently 
large population of Seyfert 2 galaxies at high redshifts to explain the observed hard X-ray background. 
Instead, deep surveys have revealed several new observational classes of X-ray sources, including
optically bright, X-ray faint galaxies (OBXFs) and X-ray bright, optically normal galaxies (XBONGs).
Neither OBXFs nor XBONGs would be classified as AGNs, based on their optical spectra.
The OBXFs are distinguished by their low X-ray--to--optical ($R$-band) flux ratios (log$(f_X/f_R)$~$<$~$-2$) 
\citep{horn01,horn03,alex03} and  are typically found at relatively low redshifts ($z~\lesssim~1$). 
Their low X-ray fluxes, combined with their quiescent optical spectra, suggest that they are 
``normal'' galaxies. Follow-up optical observations of some OBXFs have confirmed their quiescent 
character \citep{horn05}. OBXFs represent a significant fraction of the faintest 
X-ray--detected sources in deep surveys, and it remains to be seen whether some of these objects might also host AGNs. 
XBONGs, on the other hand, have large AGN luminosities ($L_{\rm X} > 10^{42}$ \lumin) in the hard (2--8~keV) X-ray band 
\citep{P3,horn05,barger05}. The X-ray brightness of these objects requires the presence of SMBH accretion activity (i.~e., 
they must be AGNs), despite their quiescent optical spectra.
There is an additional population of X-ray--detected objects with extremely faint optical counterparts that
lack classification altogether. These may include low-luminosity and optically obscured AGNs.

To account for the apparent dearth of Seyfert 2 galaxies in deep X-ray surveys, 
some of the X-ray galaxies that are optically classified as ``normal,'' 
including the OBXFs and XBONGs, may contain
AGNs that are going unrecognized because of the limited means of identifying the 
properties of distant objects from ground-based observatories. 
In particular, \citet*{moran} 
have suggested that the optical spectral characteristics of type 2 Seyferts may be 
drowned out by host galaxy light because many Seyfert 2 nuclei have low observable emission-line luminosities 
\citep[$L_{\OIII} \lesssim 10^{40}$ \lumin;][]{hao}.
This is especially likely in observations of high redshift objects, for which it is impossible to 
spatially isolate the point-like nuclear light from the extended starlight regions of galaxies.
In a sample of nearby Seyfert 2 galaxies, 
\citet{moran} found that $60\%$ of the objects would have been wrongly optically classified as normal galaxies in deep
field surveys based on their integrated spectra because starlight overwhelmed their AGN emission in the optical band.
This suggests that some of the galaxies detected in deep X-ray surveys could be mistakenly identified 
via optical spectroscopy as ``normal'' galaxies when, in fact, they harbor AGNs. 
If this is the case, then it is possible that part of the missing population of 
galaxies expected to make up the hard X-ray background is hiding among the apparently normal galaxies of deep surveys.
In the case of XBONGs, their high X-ray luminosities almost certainly require SMBH accretion activity.

As an X-ray analog to the study of \citet{moran}, and to 
further investigate the possibility that high-redshift type 2 AGNs are being detected but misidentified
in deep X-ray surveys, 
we have artificially redshifted a sample of \chandra\ observations of well-studied, nearby, Seyfert and starburst galaxies
to redshift $z = 0.3$, corresponding to a look-back time of  $\approx$$3.5$ Gyr, 
an epoch where normal/starburst galaxy X-ray luminosity functions have already
been assembled \citep{norman}. 
In \S\ 2 we describe the sample selection, data analysis, and artificial redshifting procedures. 
Section 3 presents the results of the artificial redshifting process. 
In \S\ 4 we compare the results to those of deep field surveys, and in \S\ 5 we present our conclusions.

\section{DATA ANALYSIS}
\subsection{Sample Selection and Observations}
The initial galaxy sample was selected from the public \chandra\ data archive, correlated 
with the \citet{VCV01}\footnote[4]{http://www.obs-hp.fr/www/catalogues/veron2\_10/veron2\_10.html}
catalog of quasars and active galaxies as of June 2003,
and filtered to select observations with the backside-illuminated S3 CCD of the Advanced 
CCD Imaging Spectrometer \citep[ACIS;][]{acis_ref} in which a 
$L_{2-8~\rm keV} = 5 \times 10^{39}$ \lumin\ 
point source would have a $\geq$ 7~count detection (corresponding to exposure times of 63~s 
and 106~ks for objects with photon indices
of $\Gamma = 1.4$ at redshifts of $z = 0.001$ and $z = 0.04$, respectively).
The S3~CCD is preferred because of its superior spectral resolution and enhanced response at energies below 1~keV.
These initial selection criteria yielded 
a set of 29 objects, including one starburst galaxy (NGC~2782), which appears to have been wrongly included in the catalog of \citet{VCV01}. 
Most authors refer to NGC~2782 only as a starburst \citep{NGC2782a,NGC2782c}, and \citet{NGC2782b} claim that there is no 
evidence of an AGN in this galaxy. We classify it as a non-AGN starburst. In the course of our analysis, eight sources, including 
five Seyfert 1s, were eliminated
for any of the following three reasons: having too few counts, 
having an angular size significantly exceeding that of the 8$^{\prime}$ $\times$ 8$^{\prime}$ S3 CCD, 
or showing evidence of severe data degradation by pile-up.
Pile-up occurs when two or more photons arrive on a pixel between CCD read-outs, causing loss of counts, 
and diminishing both the spatial and spectral qualities of observations (see \S\ 2.2).
One starburst galaxy (NGC 4736) and one luminous Seyfert 1/starburst galaxy (Markarian 231) 
were added to the initial sample. Our final sample consists of 23 galaxies with redshifts ranging from $z \sim 0.001$ to 0.04.
A list of the objects, including the details of their observations,
is provided in Table~\ref{tab:gal_props}. Two of the galaxies are starbursts without AGN activity, the other 
21 are Seyferts, six of which host starbursts. Ten of the galaxies also have luminous X-ray point source 
populations.

\subsection{X-ray Data Reduction}

\chandra\ ACIS data were retrieved from the \chandra\ data archive 
and reprocessed from the level 1 event files using a scripted reduction pipeline 
to ensure uniform analysis of all the observations.
This analysis employed the 
CIAO~v.3.2,\footnote[5]{Chandra Interactive Analysis of Observations (CIAO), http://cxc.harvard.edu/ciao/} 
HEASOFT~v.5.3.1,\footnote[6]{http://heasarc.gsfc.nasa.gov/ftools/, \citep{ftools}}
and IDL~v.6.0 software packages.

The ACIS data were filtered to include only data from the S3 chip. 
The CIAO tool {\tt acis\_process\_events} was used to remove the 
positional randomization that is introduced by the standard 
pipeline to avoid aliasing effects in short exposures. 
To mitigate charge transfer inefficiency (CTI), 
we employed the ACIS CTI correction software package developed by \citet{cticorr}.
The data were filtered on ``good'' ASCA grades (keeping grades 0, 2, 3, 4, and 6 to optimize the 
signal-to-background ratio) and on status. 
After examining each light curve for background flares, we accordingly adjusted 
and filtered on good time intervals to produce level 2 event files.

To avoid the adverse spatial and spectral effects of pile-up (see \S\ 2.1), we examined each 
image for read-out streaks and dark spots caused by pulse saturation, both of which are symptoms of pile-up. 
We also surveyed the literature for descriptions of 
\chandra\ data analyses of each of these sources, looking for any mention of pile-up. For those 
galaxies with evidence of pile-up, we used the PIMMS v.3.2a ACIS Pile up and Background Count Estimation 
tool\footnote[7]{http://cxc.harvard.edu/toolkit/pimms.jsp} 
to determine whether the pile-up was significant enough to affect our analysis. 
When possible, we repaired the piled-up regions of galaxies using 
observations taken with shorter frame-times, which were therefore less prone to pile-up.
Otherwise, galaxies with estimated pile-up count rates in excess of $10\%$ 
were removed from the sample.

\subsection{X-ray Artificial Redshifting}
In order to move the galaxies in our sample to $z=0.3$, we accounted for both the spatial and spectral effects 
of increased redshift on the observations. We assumed no redshift evolution.
Throughout this paper we will use the terms ``original observations'' and ``original images'' to refer to the actual 
observations of the galaxies at their true, low redshifts.
We will use ``redshifted galaxies'' and ``redshifted images'' to refer to the artificially redshifted observations
produced by the analysis described below.

We began by redshifting the energies of the detected photons, 
being mindful of the dependence 
of ACIS sensitivity on photon energy as described by
the Auxiliary Response File (ARF) associated with each \chandra\ observation.
The ARF relates the expected detector count rate 
to the incident source spectrum, accounting for the effective area 
and quantum efficiency of the telescope, filter, and detector system.
In general, the number of counts in a single \chandra\ energy band, $E_1-E_2$, is measured to be
\begin{equation}
N_{E_1-E_2} = \int_{E1}^{E2} A(E) F(E) dE,
\end{equation}
where A(E) represents the ARF and F(E) the emission spectrum of the object.
In redshifting each observation to $z=0.3$, we had to account for the 
fact that the redshifted emission spectrum from the source would encounter a non-redshifted ARF at the detector. 
Thus, to get the appropriate number of counts in the observed band $E_1-E_2$ for an artificially redshifted galaxy, 
we could not simply count the photons from the original observations in a blueshifted energy band, 
$E_{1_{orig}}(1+z)-E_{2_{orig}}(1+z)$;
the ARF also had to be effectively redshifted, i.~e.,
\begin{equation}
N_{(E_1-E_2)_{obs}} = \int_{E1_{orig}(1+z)}^{E2_{orig}(1+z)} A\left(\frac{E}{1+z}\right) F(E) dE.
\end{equation}
In practice, we accomplished this by 
multiplying the image by the ratio 
\begin{equation}
\frac{A\left(E/(1+z)\right)}{A(E)}. 
\end{equation}
Because this ratio varies significantly over the range of energies to which \chandra\ is sensitive, 
we divided the observations into narrow energy bands in which the ARF ratio varied by less than $30\%$, 
then we multiplied each narrow-band image by its corresponding ratio. 
This resulted in images in energy bands as narrow as 0.01~keV at very soft energies (below~$\sim~0.5$~keV) and bands 
as large as 1.85 keV at higher energies. 
The appropriate grating efficiency was included in creating ARFs for the original gratings observations.

After redshifting the energy in this way, we adjusted the angular extent of each narrow-band image 
to account for increased distance \citep{hogg00}, using a $\Lambda$CDM cosmology with
$\Omega_M$~=~0.3, $\Omega_{\Lambda}$~=~0.7 and $H_0$~=~70~km~s$^{-1}$~Mpc$^{-1}$. 
Because this re-sizing combined many original pixels into each redshifted pixel, 
the background level, which is usually negligible even for very long \chandra\ observations, 
was high and unrealistic in the redshifted images.
We accounted for most of the background by measuring the average background level per pixel in the original 
images, multiplying this value by an appropriate re-sizing factor and ARF ratio, 
and subtracting it from the redshifted images.

The flux of each galaxy was decreased according to its increased distance, however 
this adjustment reduced the total detectable signal from most of the sources to $\lesssim$ 10 counts.
In order to have numbers of counts comparable to those of deep field surveys, we artificially
increased the exposure time of each observation (i.~e., linearly increased the number 
of counts per pixel) such that the brightest pixel in the 0.5--8 keV band
image of each galaxy had 100 counts. (Note: This did not change the luminosities of the objects since 
we kept track of the observation times. The total artificial exposure time for each redshifted image
is listed in column 5 of Table~\ref{tab:gal_props}.) 
Even after this increase, all of the objects appeared as unresolved point sources.
We then randomized the number of counts in each pixel 
based on a Poisson distribution and blurred each image by the energy-dependent, 
on-axis \chandra\ point spread function (PSF)
to create a more realistic observation of a point source.

The narrow-band images were added up to create a final set of redshifted images in 
the following three observed-frame energy bands: full (0.5--8 keV), soft (0.5--2 keV), and hard (2--8 keV). 
Above 8.0 keV, the effective area of the Chandra mirrors rapidly declines while 
the background flux increases, reducing the signal-to-noise ratio;
the lower limit of 0.5 keV was chosen to match the well-calibrated part of the instrument response function.

\subsection{X-ray Aperture Photometry}

The CIAO tool {\tt wavdetect} was used to locate the redshifted galaxy in each full-band image. 
Aperture photometry centered at the corresponding locations was performed in the soft- and hard-band 
images, using apertures based on the $95\%$ encircled energy radius of the \chandra\ PSF \citep{memo00}.
The remaining local background in each redshifted image was measured in a surrounding region and subtracted,  
and appropriate aperture corrections were applied.
When a source was not detected in a given band, an upper limit was calculated using the Bayesian 
method of \citet*{kbn}, for $90\%$ confidence.

Hardness ratios were calculated from the hard- and soft-band counts as follows:\\
$HR$~=~$(hard~-~soft)/(hard~+~soft)$.
The hard- and soft-band fluxes and luminosities of each redshifted galaxy were estimated
based on a spectral model of the form $F(E) = N_{E_0}~E^{-\Gamma}$ photon cm$^{-2}$ s$^{-1}$~keV$^{-1}$, 
where the photon index, $\Gamma$, was determined from the measured hardness ratio ($HR = 0$ corresponds to 
$\Gamma \sim 0.5$), and the appropriate 
normalization, $N_{E_0}$, and flux in a given energy band were calculated based on the observed count rate using 
WebPIMMS.\footnote[8]{http://heasarc.gsfc.nasa.gov/Tools/w3pimms.html} The effects of Galactic neutral 
hydrogen absorption were negligible.

To avoid reporting anomalous values due to the Poisson randomization
in our redshifting procedure, and to determine the standard deviation in our values,
we iterated the redshifting process 5,000 times for each galaxy, 
calculating hardness ratio, flux, and luminosity with each iteration. 
A Gaussian curve was fit to the set of results for each quantity, and we report the resulting
mean values and standard deviations in Table~\ref{tab:lums}. 
One of the sources (NGC 4507) had very few counts in the soft band; the distributions of
0.5--2 keV flux and luminosity for this object were not Gaussian.
We report approximate average values of the soft flux and luminosity of this galaxy 
and uncertainties that span the range of possible values.

\subsection{Optical data}

Optical data were obtained from the HyperLeda database,\footnote[9]{http://leda.univ-lyon1.fr/} 
which provides total $B$-band magnitudes and $B-V$ colors of galaxies derived 
by fitting Hubble-type-dependent curves to multiple aperture photometric data \citep{leda_data}.
Because observed-frame Cousins $R$ band ($\lambda = 6700$\AA) at $z = 0.3$ 
corresponds approximately to rest-frame Johnson $V$ band ($\lambda = 5530$\AA), 
we transformed the {\em V} magnitudes from LEDA into
Cousins {\em R} magnitudes for $z=0.3$ galaxies of the appropriate Hubble types.
This transformation took into account the flux difference from the
distance as well as a k-correction from the bandpass shift.  The
k-corrections were calculated using the synthetic photometry software
package {\tt synphot}\footnote[10]{http://www.stsci.edu/resources/software\_hardware/stsdas/synphot/SynphotManual.pdf}
and a galaxy spectral template matched in Hubble type to each object; k-correction values ranged from
$-0.2$ to $-0.5$.

$B-V$ colors were not available from LEDA for four of the galaxies in our sample 
(NGC~424, NGC~2110, MCG-05-23-016, and IC~2560).
For these objects, the colors typical of their various Hubble types were used \citep{fuku95}.

\section{REDSHIFTED GALAXY PROPERTIES}
The practice adopted here of calculating X-ray luminosity from a simple spectral model, namely a hardness ratio 
and a photon count rate, is standard among deep field observations where low  numbers of detected 
X-ray photons prevent a more complicated analysis.
We find that the simply measured 2--8 keV luminosities of our artificially redshifted observations are 
consistent within a factor of $\approx$1--3 with the luminosities determined by much more sophisticated 
spectral modeling, as available in the literature (see references in Table~\ref{tab:gal_props}).

The exposure time necessary to detect 100 counts in the brightest pixel of each full-band image 
(before blurring) is reported in columns 4 and 5 of Table~\ref{tab:gal_props}. Clearly, some of these exposure
times are unreasonably long. For comparison, we list the number 
of counts expected from a 1~Ms \chandra\ ACIS-S observation of each of these objects at $z = 0.3$ in columns 
6 and 7 of Table~\ref{tab:gal_props}. Estimating that 10 counts are required for a full-band detection, 
we find that three sources (NGC~4395, NGC~4736, and Circinus) would not be detected near the aimpoint of a 
1~Ms \chandra\ ACIS-S observation.

\begin{figure*}[t!]
\centerline{
\psfig{figure=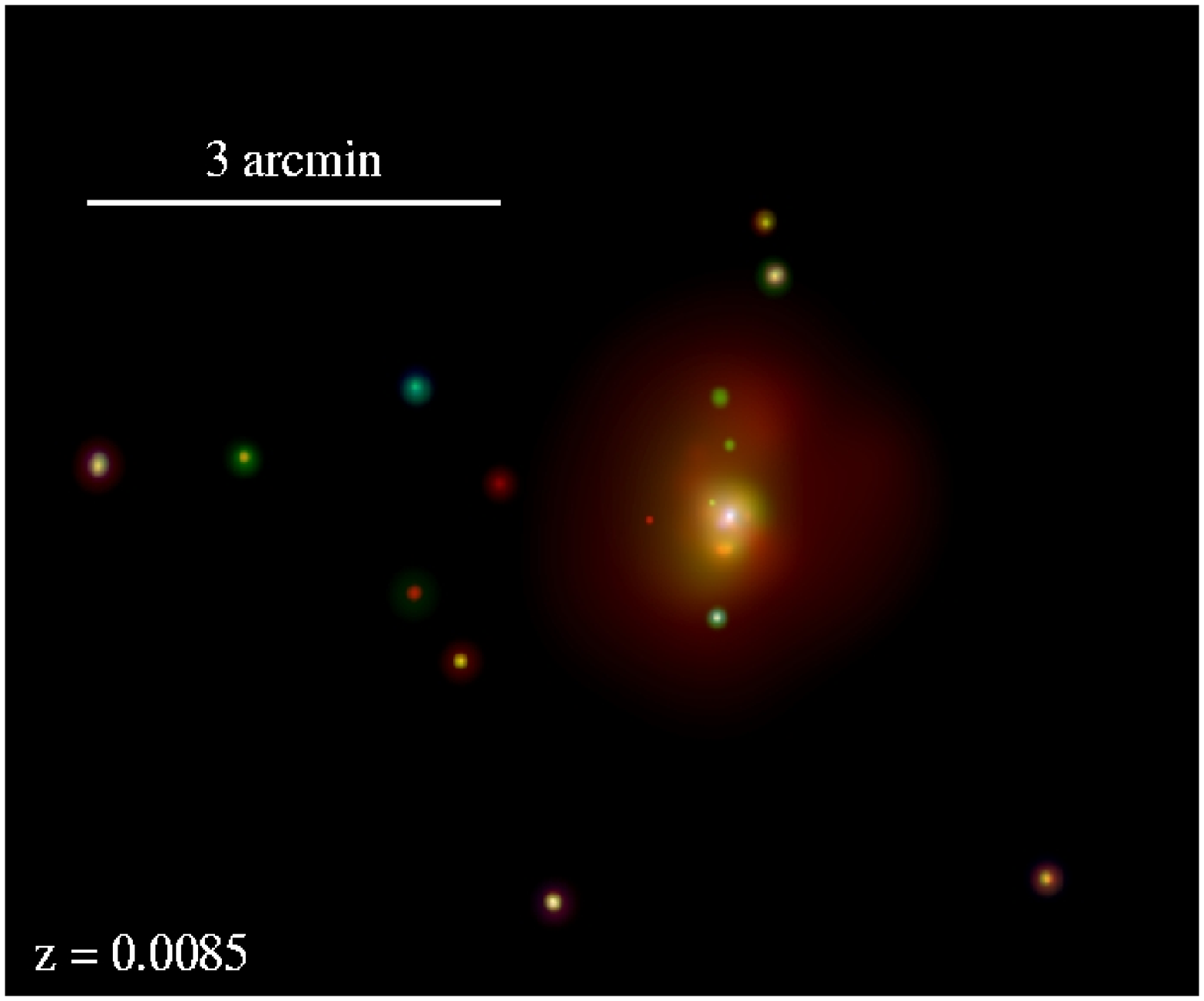,width=5 in}
\psfig{figure=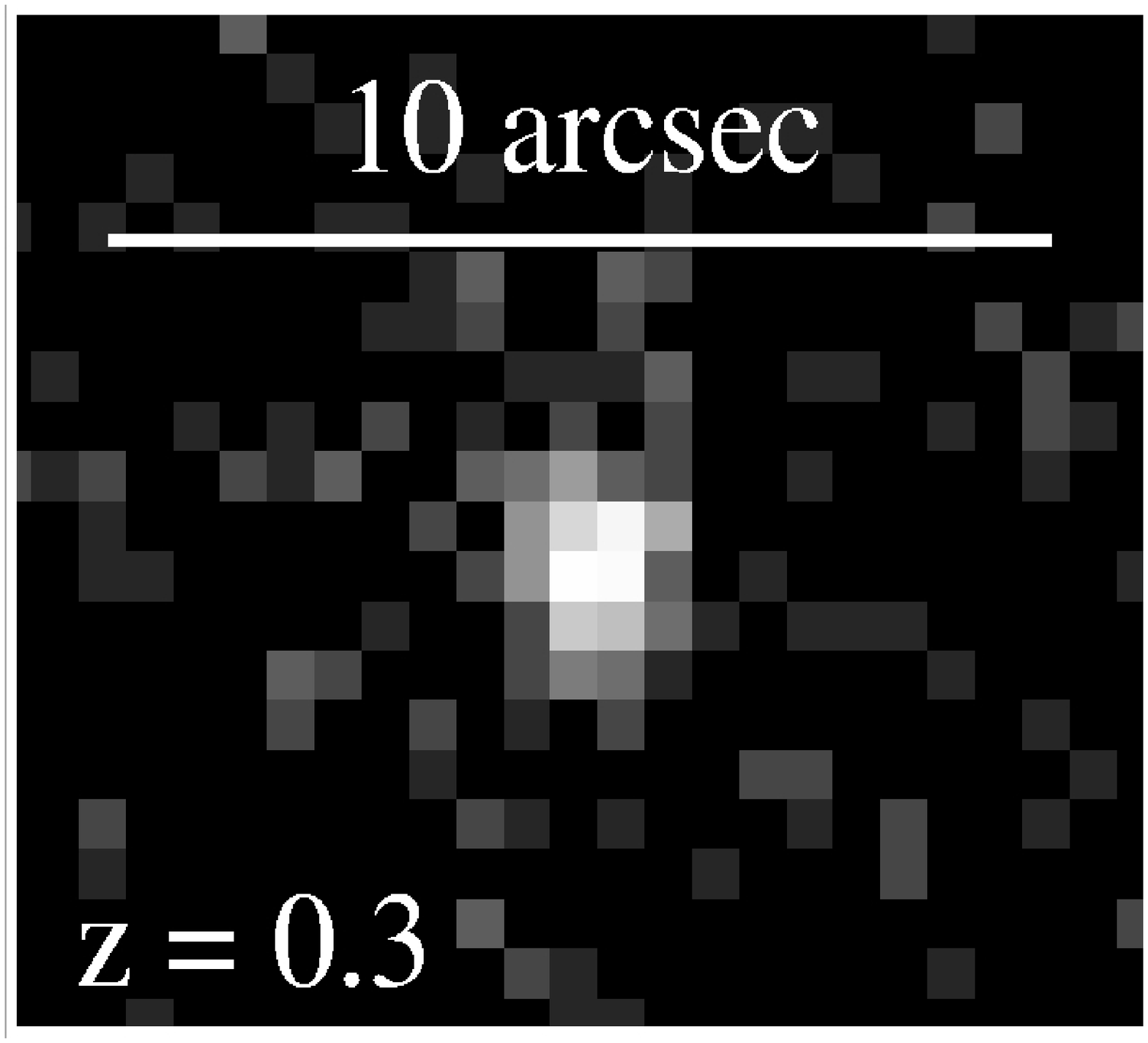,width=2 in}
}
\caption{
{Left: Adaptively smoothed, \chandra\ ACIS-S, three-color image of starburst galaxy NGC~2782. Red is 0.35--1.2~keV, green is 1.2--3.0~keV,
and blue is 3.0--8.0~keV. Right: Artificially redshifted image of NGC~2782 in 0.5--8.0~keV energy band.
}
\label{fig:image}
}
\end{figure*}

A 3-color X-ray image of the starburst galaxy NGC~2782 at its actual, low redshift is presented in Figure~\ref{fig:image}. 
The colors in the image correspond to energy bands of 0.35--1.2~keV (red), 1.2--3.0~keV (green), and 3.0--8.0~keV (blue). 
Evident in this image are the diffuse soft X-ray emission surrounding the central starburst and a collection of point sources of various colors. 
We also present a grayscale artificial image of the same starburst galaxy at $z = 0.3$ in the full (0.5--8.0~keV) energy band. 
At this distance, the galaxy's structure is unresolvable, and it appears as a point source. This is typical of the redshifted 
observations of this study.

\begin{figure*}[t!]
\centerline{
\plotone{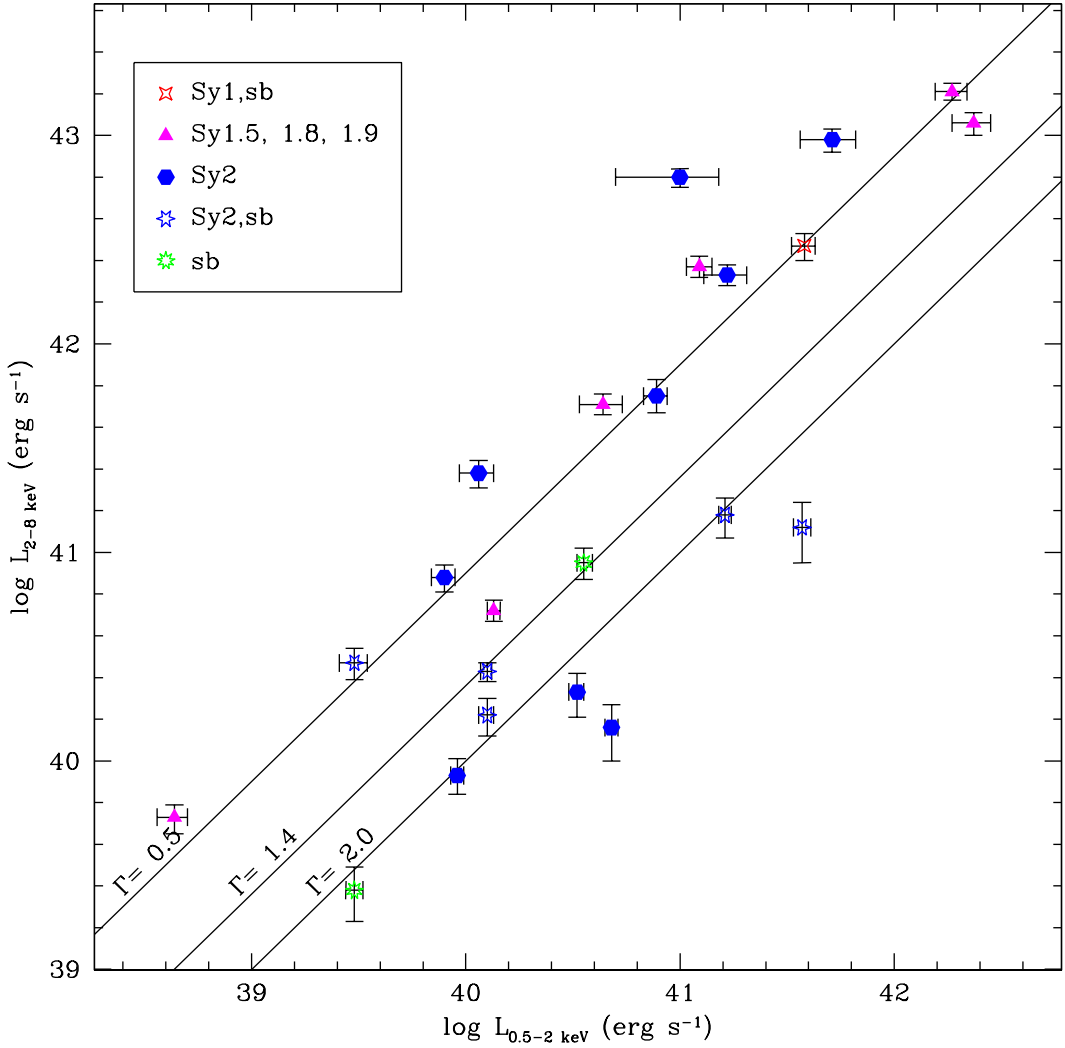}
}
\caption{
{Hard (2--8 keV) versus soft (0.5--2 keV) luminosities of the 
artificially redshifted galaxies of our sample.
The solid blue hexagons signify Seyfert 2 galaxies; open, blue, six-pointed 
stars are Seyfert 2 galaxies with starbursts;
solid magenta triangles are Seyfert 1.5, 1.8, and 1.9 galaxies; 
open, red, 4-pointed stars are Seyfert 1 galaxies with starbursts; 
and open, green, nine-pointed stars are starbursts without evidence of AGNs.
The solid lines follow photon indices $\Gamma = 0.5$, $\Gamma = 1.4$ (characteristic of the X-ray background),
and $\Gamma = 2.0$.
}
\label{fig:color}
}
\end{figure*}

In Figure~\ref{fig:color} we plot the hard- and soft-band luminosities of the galaxies in our sample. 
We find that the Seyfert~1 galaxy is among the most luminous 
of the sample, which is not surprising since Seyfert~2 
galaxies suffer from more X-ray absorption than Seyfert~1s.
This effect would have been more pronounced had not 
several X-ray bright Seyfert 1 galaxies been eliminated from the sample due to pile-up.

In Figure~\ref{fig:histo} we present a histogram of the hardness ratios of the galaxies in our sample,
separated based on the presence or absence of starbursts. 
We find that non-starburst AGNs cover a wide range of hardness ratios but are predominantly hard ($HR$ $>$ 0).
This trend toward harder spectra among AGNs is not surprising; SMBH accretion produces a strong non-thermal 
continuum that is bright across the entire X-ray spectrum. In particular, Seyfert~2s (signified by black dots in 
Figure~\ref{fig:histo}) are often highly absorbed at low X-ray energies and therefore have hard spectra.
However, we find that about half of the Seyfert~2s in this sample are soft ($HR$ $<$ 0).
In the case of galaxies hosting both an AGN and a starburst, the low hardness ratio is likely due to 
starburst emission affecting the shape of the spectrum \citep[e.g.,][]{lev01}. Starbursts contribute significantly 
to the soft end of the X-ray spectrum, with a thermal component from supernova remnants and O and B star winds, in addition to 
radiation from  X-ray binaries that may be soft or hard, depending on the properties of the accreting systems. 
Both of the pure starburst galaxies in this sample have $HR$ $<$ 0.
The three non-starburst Seyfert 2s in this sample with the softest spectra, NGC~4374, NGC~4552, and NGC~5194, all have
significant X-ray point source populations and luminous, extended, soft X-ray emission that affect 
their integrated spectral properties \citep{NGC4374b,NGC4552,NGC5194}. In addition, NGC~4552 and NGC~5194 are known 
to be very weak AGNs, so non-nuclear X-ray sources may dominate their X-ray emission across the \chandra\ bandpass.
It is also possible that some of the optically obscured (i.e., type 2) AGNs in this sample appear X-ray soft because they are
not heavily absorbed at X-ray energies.

\section{COMPARISON WITH DEEP FIELD SURVEYS}
\subsection{Deep Field Survey Data}
We compare our results with those of three X-ray surveys: 
the \chandra\ Deep Field North survey \citep[CDF-N;][]{alex03}, the ROSAT Ultra Deep Survey 
\citep[UDS;][]{lehmann}, and the \chandra\ Multiwavelength Project \citep[ChaMP;][]{champ}.
Together these three surveys sample over 4 orders of magnitude in X-ray flux, 
allowing comparison with a broad range of X-ray source types.

\begin{figure*}[t!]
\centerline{
\plotone{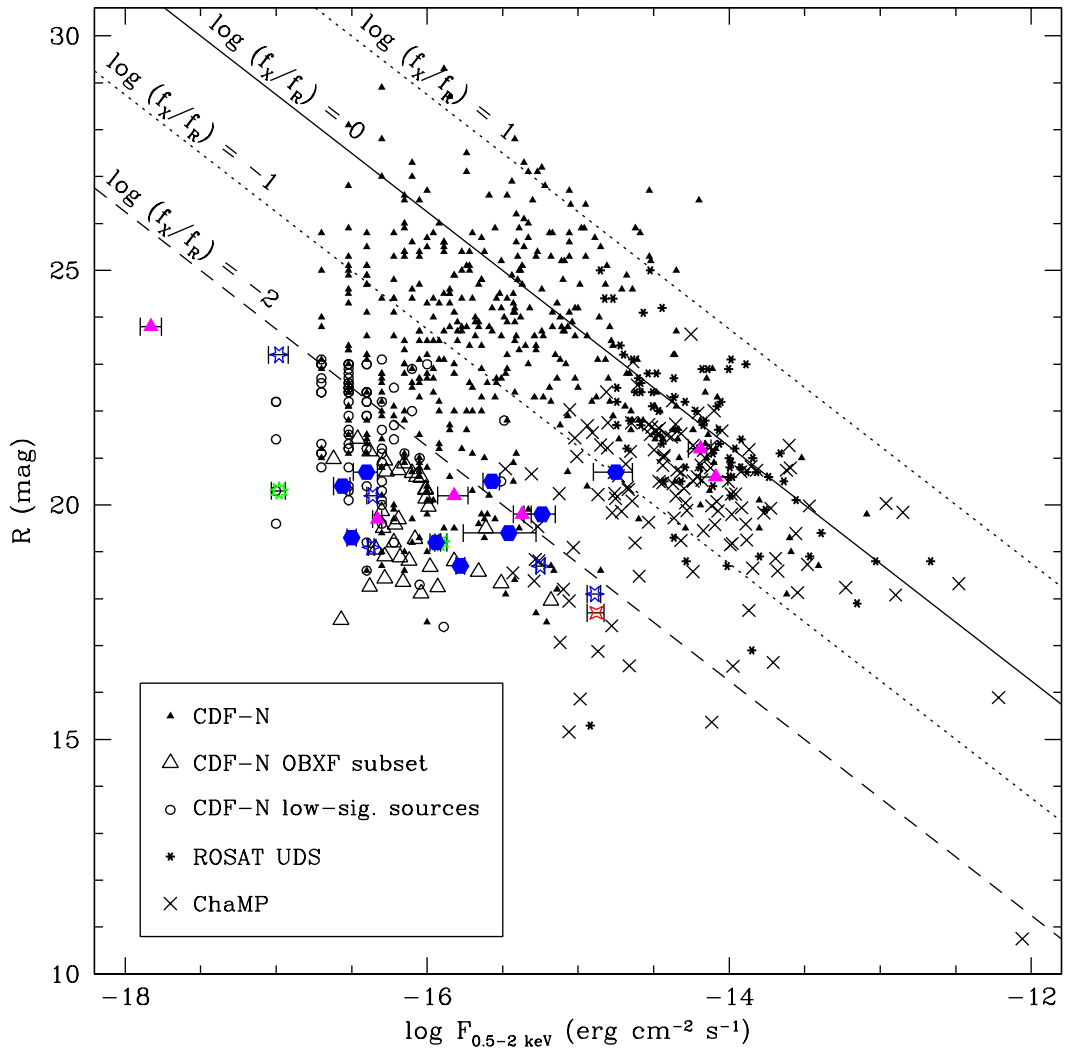}
}
\caption{
{Optical $R$-band magnitude versus 0.5--2 keV X-ray flux. Colored symbols represent artificially redshifted 
galaxies as in Fig.~\ref{fig:color}.
The main CDF-N objects are small triangles, the subset of OBXF objects are large, open triangles, and
the supplementary, lower-significance sources are open circles. Small black stars represent ROSAT UDS data, 
and large \textsf{X}'s represent ChaMP data. 
The slanted lines follow constant values of log$(f_X/f_R)$ as follows:
the solid line marks log$(f_X/f_R)$~=~0, 
the dotted lines mark log$(f_X/f_R)$~=~$\pm$1, and the dashed line marks  log$(f_X/f_R)$~=~$-2$.
The region below log~$(f_X/f_R)$~=~$-2$ is believed to be typically populated by quiescent galaxies.}
\label{fig:rmag}
}
\end{figure*}

The CDF-N survey comprises a 2~Ms \chandra\ observation of a 448~arcmin$^{2}$ field, reaching on-axis
flux limits in the 0.5--2 keV and 2--8 keV bands of
2.5~$\times$~$10^{-17}$ and 1.4~$\times$~$10^{-16}$ \flux, respectively.
The main CDF-N catalog contains 503 X-ray-detected sources \citep{alex03}, for all of which optical 
counterparts were identified using
the 8.2-meter Subaru telescope \citep{barger03}. Follow-up optical spectroscopy was performed on an 
OBXF (log$(f_X/f_R)$~$\lesssim$~$-2$) subset of these sources \citep{horn03}.
This subset was revealed to consist primarily of apparently normal galaxies, based on optical identification.
The median redshift of the subset was $z \approx 0.3$, making it quite comparable 
to the redshifted galaxy sample of this paper.
A supplementary CDF-N catalog of 79 lower significance X-ray sources was created by searching for X-ray counterparts 
to optically identified sources; these were therefore also OBXF galaxies.
In Figures~\ref{fig:cdfcolor} and \ref{fig:rmag} we plot the main CDF-N catalog objects 
as small, solid, black triangles, the OBXF subset as large open triangles, 
and the lower-significance supplementary objects as open circles.

The ROSAT UDS consists of three independent samples of sources observed in the direction of the 
Lockman Hole with the 
ROSAT PSPC and/or HRI for 207~ks (PSPC) and 1112~ks (HRI), reaching a flux limit of 1.2 $\times$ $10^{-15}$ \flux\ 
in the 0.5--2~keV energy band.
Ninety-four X-ray sources were identified in the survey \citep{lehmann}, 
and Cousins $R$-band optical magnitudes were obtained with the
Low Resolution Imaging Spectrometer on the Keck I and II 10-meter telescopes for most of the sources.
More than half of the UDS sources are type I AGNs, and only one definitive
``normal'' galaxy was identified. 
In Figure~\ref{fig:rmag}, we plot 77 of the sources that have reliable $R$-band magnitudes as small black stars for 
comparison with the results of this project.

The ChaMP data used for comparison in Figure~\ref{fig:rmag} include 
125 serendipitous X-ray sources detected in six \chandra\ observations \citep{champ}.
The observations had exposure times ranging from 29.1 to 114.6~ks, and the serendipitous 
source 0.5--2~keV fluxes range from 3.3~$\times$~$10^{-16}$ to 8.7~$\times$~$10^{-13}$ \flux.
Follow-up optical imaging observations of the sources were performed with the 
NOAO 4-meter telescope Mosaic CCD cameras, using an SDSS $r'$ filter. 
Based on optical spectroscopic observations of the ChaMP objects used for comparison in this paper, 
50\% of them are type 1 AGNs, 22\% are narrow emission-line 
galaxies (including starbursts and AGNs), and 18\% are absorption-line galaxies. In Figure~\ref{fig:rmag} 
we plot the SDSS $r'$ magnitudes versus the soft X-ray fluxes 
of each of these sources as large \textsf{X}s.

\subsection{Discussion of Deep Field Survey Comparison}
When optical spectroscopy is not available, source classification in deep field surveys
is generally carried out using optical--to--X-ray flux ratios, hardness
ratios, and X-ray luminosity. The sample we have assembled may
provide a useful prior for source classification in such studies. As
we have effectively illustrated in \S~3, a significant number of ``pure''
Seyfert~2s (roughly one-third of our sample) have $HR$~$<$~0.   Notably,
$HR$~$<$~0 was one of the selection criteria applied to the \citet{norman} 
\chandra\ Deep Field sample to separate normal galaxies from AGNs.
Apparently, using simple combinations of parameters (such as $L_x$ and
$HR$; see Figure~\ref{fig:color}) cannot resolve this classification issue.

\begin{figure*}[t!]
\centerline{
\plotone{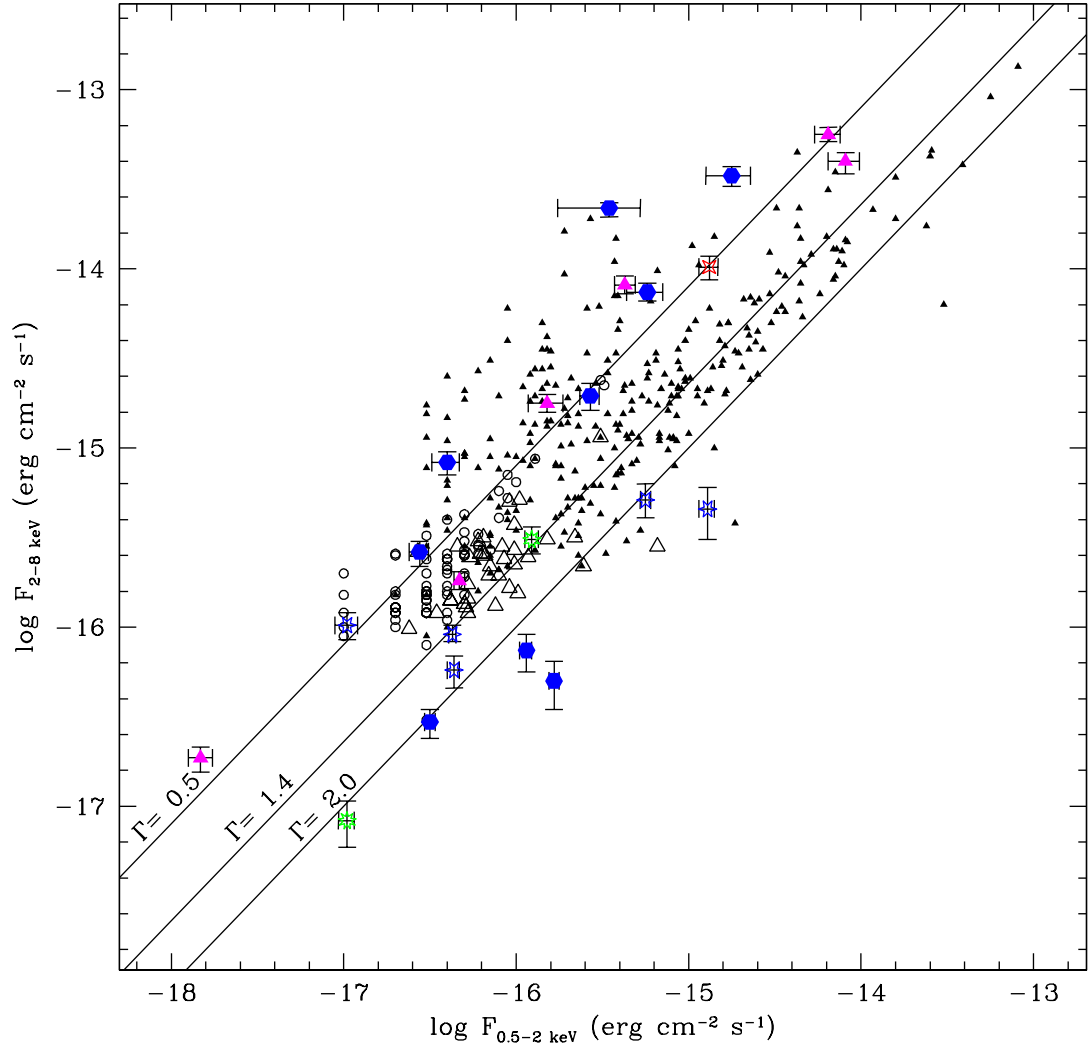}
}
\caption{
{Hard versus soft X-ray flux.  Colored symbols represent artificially redshifted 
galaxies as in Fig.~\ref{fig:color}.
The main CDF-N objects are small triangles, the subset of OBXF objects are large, open triangles, and
 the supplementary, lower-significance sources are open circles.
The solid lines indicate photon indices of $\Gamma = 0.5$, $\Gamma = 1.4$ (characteristic of the X-ray background),
and $\Gamma = 2.0$.}
\label{fig:cdfcolor}
}
\end{figure*}

The artificially redshifted galaxies of our sample span almost the entire range of soft and hard X-ray 
fluxes observed in the CDF-N,
as can be seen in Figure~\ref{fig:cdfcolor}, and all but one (NGC 4395) are within the flux limits of that survey.
The solid lines in Figure~\ref{fig:cdfcolor} indicate the hard- to soft-band flux 
relation of various power-law models, including $\Gamma$~=~1.4, which is characteristic of the X-ray background.
We find that the CDF-N sources occupy a somewhat narrower 
range in hard-band flux per unit soft-band flux than the galaxies of our sample, clustering 
around $\Gamma$~=~1.4. This is expected not only because $\Gamma$~=~1.4 is a typical photon index,
but also because the deep survey assumes a $\Gamma$~=~1.4 power law 
for objects with low numbers of detected X-ray photons. (Both the 
X-ray faint objects of the CDF-N OBXF subset and the lower-significance 
supplementary catalog objects were assigned a photon index of 
$\Gamma$~=~2.0, a value more typical of normal galaxies, in cases of low number counts.)
The spread in X-ray color of the redshifted galaxy population 
demonstrates that the spectral shape of a galaxy's integrated X-ray emission 
is not necessarily indicative of its nature. For example, we find that obscured AGNs with significant star formation can appear quite soft, 
as discussed in \S~3.

Figure~\ref{fig:rmag} plots optical--to--X-ray flux ratios 
of the redshifted galaxies in our sample and the deep field objects. These reveal
that our redshifted Seyfert galaxies are significantly brighter at $R$ band than the majority of the
deep field X-ray sources. This may in part reflect the bias in our sample, which, being selected 
on the basis of availability in the \chandra\ archive, is likely to favor well-known, optically bright galaxies. 
The slanted lines on this plot follow constant log$(f_X/f_R)$ values, where 
log$(f_X/f_R)$~=~log$(f_X)$~+~5.50~+~$R/2.5$, dividing
the plot into regions of assumed typical X-ray--to--optical 
flux ratios for different source classes.
According to the results of X-ray surveys in which source classification has been based on 
follow-up optical spectroscopy (and is therefore subject to the limitations of spectroscopic identification, as 
discussed in \citet{moran} and in \S~1 of this paper),
AGNs generally fall in the region \mbox{$-1$~$<$~log$(f_X/f_R)$~$<$~1}, while
quiescent galaxies fall in the region log$(f_X/f_R)$~$<$~$-2$.
The transition zone of $-2$~$<$~log$(f_X/f_R)$~$<$~$-1$ is
populated by both AGNs and starburst galaxies \citep{emss,rosat,asca,horn01}. 
The galaxies of our sample are noticeably optically bright relative to their X-ray fluxes;
more than $85\%$ of them fall below the log$(f_X/f_R)$~=~$-1$ AGN cut-off, and $\approx$$70\%$
reside below log$(f_X/f_R)$~=~$-2$, in the realm of normal galaxies. All but one of the Seyfert/starburst
sources have log$(f_X/f_R)$~$<$~$-2$, as do both of the pure starbursts.
The location of most of the redshifted objects in the realm of quiescent and starburst galaxies (log$(f_X/f_R)$~$<$~$-2$)
implies that these galaxies would not be recognized as AGNs if observed at $z = 0.3$. 
The relatively low X-ray luminosities of the majority of the sources ($L_{0.5-2~\rm keV} \sim 10^{40}$ \lumin) also would not suggest
AGN presence. This sample is most consistent with the OBXF CDF-N sources, many of which have been identified as normal galaxies.

\section{CONCLUSIONS}
\begin{figure*}[t!]
\centerline{
\plotone{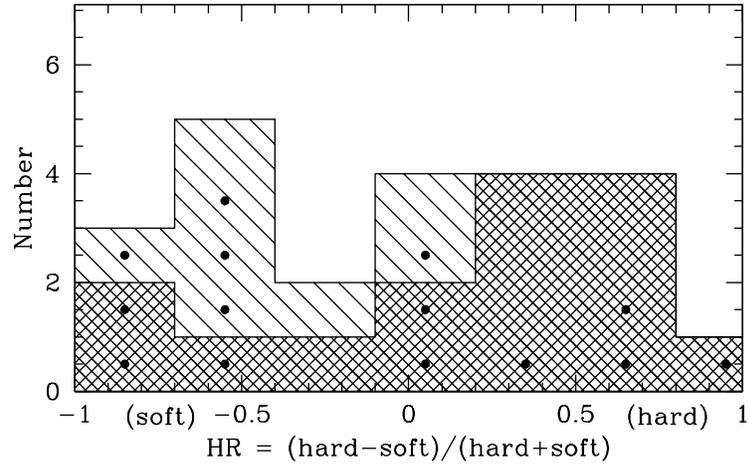}
}
\caption{
{Histogram of redshifted galaxy hardness ratios, 
separated based on presence or absence of star formation activity. 
The closely spaced, crossed hatches are galaxies with only AGNs; the widely spaced 
hatches are galaxies with starbursts (six of these also host Seyfert nuclei). Seyfert 2 galaxies 
are marked with black dots.}
\label{fig:histo}
}
\end{figure*}

We find that, when artificially redshifted to $z = 0.3$, a sample of nearby AGNs has integrated 
X-ray properties and X-ray--to--optical flux ratios consistent with the ``normal'' galaxies of deep X-ray surveys. 
In addition, we see evidence that non-AGN emission can significantly affect the shapes of the X-ray spectra 
of galaxies, as demonstrated by the soft spectra of the AGNs hosting starbursts in our sample
(Figure~\ref{fig:histo}). Distinguishing obscured AGNs from the high-redshift X-ray detected galaxies 
whose X-ray emission is purely powered by X-ray binaries and hot gas from starbursts will require 
more sophisticated diagnostics than simple hardness ratios, X-ray--to--optical flux ratios, and X-ray luminosities. 
We also find that longer X-ray observations ($\sim$3--10~Ms) will be necessary to detect some of these interesting objects, 
including starbursts and low-luminosity Seyfert galaxies, at $z = 0.3$.

Recognizing that the redshifted galaxy sample presented here is biased, 
and that objects such as these with $R < 23$ are relatively rare in deep field observations, 
we make no claims about the overall population of deep field objects. 
However, this work does offer X-ray support to the optical finding of \citet{moran} that type 2 
AGNs could easily go unrecognized in observations of distant galaxies. 
This result has several possible implications:

1. Some fraction of the missing population of type 2 AGNs thought to comprise the unresolved hard X-ray background 
could be hidden among the so-called normal, OBXF galaxies of deep X-ray surveys. 
Seyfert-luminosity AGNs such as the ones studied in this paper can be found at X-ray--to--optical 
flux ratios $<$~$10^{-2}$, imitating quiescent and starburst galaxies. Their X-ray spectra 
may be dominated by star formation at the energies to which \chandra\ and ROSAT are sensitive, and therefore
the hardness ratio may not indicate the level of obscuration or the $>$~10~keV brightness of these objects.

2. X-ray selected studies of galaxies which use criteria such as
X-ray luminosity, hardness ratio and X-ray--to--optical flux ratio 
to segregate normal galaxies and AGNs \citep[e.g.,][]{norman} 
could have higher levels of AGN contamination than previously estimated.
Therefore, studies of SMBH accretion may be misled in their analyses of 
AGN evolution by the incomplete identification of X-ray detected sources. 
These observational limitations must be worked out before the details 
of AGN evolution can be well understood.

Mrk 231, the only Seyfert~1 galaxy included in our sample, provides an
interesting case study in potential identification problems with low
signal-to-noise ratio X-ray survey data.  With broad optical emission
lines, $L_{\rm 2-8~keV}\sim10^{42}$ \lumin, and a moderately hard
$HR=0.07$, this source would likely be identified as an X-ray absorbed
Seyfert~1 galaxy at $z\sim0.3$.  However, the complex 0.5--8.0~keV X-ray
spectrum includes soft starburst emission, and the direct X-ray continuum
is blocked by a Compton-thick absorber \citep{braito+04}.  Extrapolating
simply from the simulated 0.5-8.0~keV \chandra\ survey data would lead to
a gross underestimate of the hard X-ray luminosity of this source.

These results will benefit from more careful study with a larger, well-defined 
sample of nearby galaxies. Overlap with the sample of \citet{moran}, and an extension 
of their optical spectroscopy study would also be valuable.

\acknowledgements
We thank J.~C.~McDowell for valuable advice regarding the artificial redshifting process. 
This work was made possible by \chandra\ X-ray Center grant G03-4137. 
Support for SCG was provided by NASA through the {\em Spitzer} Fellowship
Program, under award 1256317. This work made use of images and/or data products provided 
by the Chandra Multiwavelength Project (ChaMP; Green et al. 2002), supported by NASA. 
Optical data for the ChaMP are obtained in part through the National Optical Astronomy Observatory (NOAO), 
operated by the Association of Universities for Research in Astronomy, Inc. (AURA), 
under cooperative agreement with the National Science Foundation.
This research has also made use of the NASA/IPAC Extragalactic Database (NED) which is operated by the 
Jet Propulsion Laboratory, California Institute of Technology, under contract with the
 National Aeronautics and Space Administration.


\clearpage
\begin{landscape}
\begin{deluxetable*}{lccrrrrrccc}
\tablecolumns{11}
\tablewidth{0pt}
\tablecaption{Observations
\label{tab:gal_props}
}
\tabletypesize{\scriptsize}
\tablehead{
\colhead{Galaxy\tablenotemark{a}} &
\colhead{Obs. ID} &
\colhead{Start Date} &
\multicolumn{2}{c}{Exposure time (ks)} &
\colhead{} &
\multicolumn{2}{c}{Counts per 1 Ms exposure\tablenotemark{b}} &
\colhead{Grating}  & \colhead{Type\tablenotemark{c}} & \colhead{References} \\
\cline{4-5} 
\cline{7-8} 
\colhead{} & \colhead{} & \colhead{} & 
\colhead{Actual} & \colhead{Artificial} &
\colhead{} &
\colhead{0.5--2 keV} & \colhead{2--8 keV} &
\colhead{} &  \colhead{} & \colhead{}
}
\startdata
NGC 424           & 3146  & 2002 Feb 04  &  9.2\phantom{00}     &      811.9\phantom{00} &  &   71\phantom{000} &     78\phantom{0000}    & NONE  &  Sy2 & 1       \\
NGC 1068*          & 344   & 2000 Feb 21 & 47.4\phantom{00}   &      328.3\phantom{00} &  &  344\phantom{000} &     30\phantom{0000}    & NONE  &  Sy2,sb & 2      \\
NGC 2110*          & 3418  & 2001 Dec 20 & 46.8\phantom{00}   &      245.0\phantom{00} &  &   45\phantom{000} &    392\phantom{0000}     & HETG  &  Sy2 & 3        \\
MARKARIAN 3       & 873   & 2000 Mar 18  &100.6\phantom{00}    &     1526.5\phantom{00} &  &   15\phantom{000} &     93\phantom{0000}    & HETG  &  Sy2  & 4       \\
NGC 2782          & 3014  & 2002 May 17  & 29.6\phantom{00}    &     4949.9\phantom{00} &  &   33\phantom{000} &     15\phantom{0000}    & NONE  & sb & 5          \\
MCG-05-23-016*     & 2121  & 2000 Nov 14  &51.9\phantom{00}    &      225.8\phantom{00} &  &  168\phantom{000} &    740\phantom{0000}     & HETG  &  Sy1.9 & 6     \\
NGC 3079          & 2038  & 2001 Mar 07  & 26.6\phantom{00}    &    17846.2\phantom{00} &  &   12\phantom{000} &      3\phantom{0000}    & NONE  &  Sy2,sb & 7     \\
IC 2560            & 1592  & 2000 Oct 29  & 9.8\phantom{00}    &    11084.2\phantom{00} &  &    7\phantom{000} &     10\phantom{0000}     & NONE  &  Sy2 & 8       \\
NGC 3227          & 860   & 1999 Dec 30  & 46.4\phantom{00}    &     5527.9\phantom{00} &  &    4\phantom{000} &     22\phantom{0000}    & HETG  &  Sy1.5 & 9      \\
NGC 4151*          & 348   & 2000 Mar 07  &22.9\phantom{00}    &      536.0\phantom{00} &  &  112\phantom{000} &    267\phantom{0000}     & NONE  &  Sy1.5 &10     \\
NGC 4258*          & 1618  & 2001 May 28  &20.9\phantom{00}    &    16769.3\phantom{00} &  &   13\phantom{000} &      8\phantom{0000}     & NONE  &  Sy1.9 &11     \\
NGC 4303          & 2149  & 2001 Aug 07  & 21.8\phantom{00}    &    47450.4\phantom{00} &  &   12\phantom{000} &      5\phantom{0000}    & NONE  &  Sy2,sb & 12    \\
NGC 4374          & 803   & 2000 May 19  & 28.5\phantom{00}    &     6943.3\phantom{00} &  &   44\phantom{000} &      3\phantom{0000}    & NONE  &  Sy2 & 13       \\
NGC 4388           & 1619  & 2001 Jun 08  &20.0\phantom{00}    &     3516.1\phantom{00} &  &   11\phantom{000} &     26\phantom{0000}     & NONE  &  Sy2 & 14      \\
NGC 4395          & 882   & 2000 Jun 20  & 16.7\phantom{00}    &   113329.9\phantom{00} &  &    0\phantom{000} &      1\phantom{0000}    & NONE  &  Sy1.8 & 15     \\
NGC 4507          & 2150  & 2001 Mar 15  &138.0\phantom{00}    &      764.4\phantom{00} &  &    9\phantom{000} &    225\phantom{0000}    & HETG  &  Sy2 & 16       \\
NGC 4552          & 2072  & 2001 Apr 22  & 54.4\phantom{00}    &     6818.5\phantom{00} &  &   31\phantom{000} &      5\phantom{0000}    & NONE  &  Sy2 & 17       \\
NGC 4736          & 808   & 2000 May 13  & 47.4\phantom{00}    &    35075.7\phantom{00} &  &    3\phantom{000} &      0\phantom{0000}    & NONE  &  sb & 18        \\
MARKARIAN 231     & 4029  & 2003 Feb 11  & 38.6\phantom{00}    &      202.2\phantom{00} &  &  351\phantom{000} &    401\phantom{0000}    & NONE  &  Sy1,sb & 19    \\
NGC 5135          & 2187  & 2001 Sep 04  & 26.4\phantom{00}    &     1353.2\phantom{00} &  &  151\phantom{000} &     30\phantom{0000}    & NONE  &  Sy2,sb & 20    \\
NGC 5194          & 1622  & 2001 Jun 23  & 26.8\phantom{00}    &    25786.0\phantom{00} &  &    9\phantom{000} &      2\phantom{0000}    & NONE  &  Sy2 & 21       \\
CIRCINUS GALAXY*  & 356   & 2000 Mar 14  & 23.1\phantom{00}    &    15901.0\phantom{00} &  &    3\phantom{000} &      4\phantom{0000}     & NONE  &  Sy2,sb & 22   \\
NGC 5506*          & 1598  & 2000 Dec 31  &88.9\phantom{00}    &      137.6\phantom{00} &  &  203\phantom{000} &    560\phantom{0000}     & HETG  &  Sy1.9 & 23    \\
\enddata	
\tablenotetext{a}{Asterisk indicates piled up observation repaired with shorter frame-time observation.}
\tablenotetext{b}{Assumes on-axis observations with the \chandra\ ACIS-S3 chip.}
\tablenotetext{c}{Indicates Seyfert classification (see references, column 8). Starbursts denoted by ``sb''.}
\tablerefs{{\tiny{References: 
(1) \citealt{NGC424}; 
(2) \citealt{NGC1068a,NGC1068b}; 
(3) \citealt{NGC2110}; 
(4) \citealt{MKN3}; 
(5) \citealt{NGC2782a,NGC2782b}; 
(6) \citealt{MCG}; 
(7) \citealt{NGC3079a,NGC3079b}; 
(8) \citealt{IC2560}; 
(9) \citealt{NGC3227}; 
(10) \citealt{NGC4151a,NGC4151b}; 
(11) \citealt{NGC4258}; 
(12) \citealt{NGC4303a,NGC4303b}; 
(13) \citealt{NGC4374a,NGC4374b};
(14) \citealt{NGC4388}; 
(15) \citealt{NGC4395a,NGC4395b}; 
(16) \citealt{NGC4507}; 
(17) \citealt{NGC4552,NGC4552ps};
(18) \citealt{NGC4736};
(19) \citealt{MKN231a,MKN231b}; 
(20) \citealt{NGC5135}; 
(21) \citealt{NGC5194}; 
(22) \citealt{CIRCINUSa,CIRCINUSb}; 
(23) \citealt{NGC5506a,NGC5506b}}}
}
\end{deluxetable*}
\clearpage
\end{landscape}
\begin{deluxetable*}{lcccccc}
\tablecolumns{7}
\tablewidth{0pt}
\tablecaption{Redshifted Galaxy Properties
\label{tab:lums}
}
\tabletypesize{\scriptsize}

\tablehead{
\colhead{} &
\colhead{HR} &
\colhead{$F_{0.5-2~\rm keV}$} &
\colhead{$F_{2-8~\rm keV}$} &
\colhead{$L_{0.5-2~\rm keV}$} &
\colhead{$L_{2-8~\rm keV}$}\\
\colhead{Galaxy} &
\colhead{$(h-s)/(h+s)$} &
\colhead{($10^{-16}$ \flux)} &
\colhead{($10^{-16}$ \flux)} &
\colhead{($10^{40}$ \lumin)} &
\colhead{($10^{40}$ \lumin)} 
}
\startdata
NGC 424             &  $ 0.04\pm0.10$  &  $ 2.7  \pm0.4  $  &  $20    \pm4\phantom{2}    $  &  $ 7.7  \pm1.0  $  &  $57    \pm10    $   \\
NGC 1068            &  $-0.83\pm0.04\phantom{-}$  &  $13    \pm1\phantom{2}    $  &  $ 4.5  \pm1.5  $  &  $37    \pm4\phantom{2}    $  &  $13    \pm4\phantom{1}    $   \\
NGC 2110            &  $ 0.79\pm0.05$  &  $18    \pm5\phantom{2}    $  &  $330    \pm40\phantom{2}  $  &  $51    \pm15   $  &  $950    \pm120  $   \\
MARKARIAN 3         &  $ 0.73\pm0.06$  &  $ 5.7  \pm1.3  $  &  $75    \pm8\phantom{2}    $  &  $16    \pm4\phantom{1}    $  &  $220    \pm20\phantom{2}    $   \\
NGC 2782            &  $-0.38\pm0.07\phantom{-}$  &  $ 1.2  \pm0.1  $  &  $ 3.1  \pm0.5  $  &  $ 3.6  \pm0.3  $  &  $ 8.9  \pm1.5  $   \\
MCG-05-23-016       &  $ 0.63\pm0.06$  &  $65    \pm11   $  &  $560    \pm50\phantom{2}  $  &  $190    \pm30\phantom{1}  $  &  $1600    \pm200\phantom{1}    $   \\
NGC 3079            &  $-0.58\pm0.06\phantom{-}$  &  $ 0.43 \pm0.03 $  &  $ 0.57 \pm0.12 $  &  $ 1.3  \pm0.1  $  &  $ 1.6  \pm0.3  $   \\
IC 2560             &  $ 0.14\pm0.08$  &  $ 0.27 \pm0.03 $  &  $ 2.6  \pm0.4  $  &  $ 0.79 \pm0.10 $  &  $ 7.6  \pm1.2  $   \\
NGC 3227            &  $ 0.71\pm0.06$  &  $ 1.5  \pm0.3  $  &  $18    \pm2\phantom{2}    $  &  $ 4.4  \pm1.0  $  &  $51    \pm6\phantom{1}    $   \\
NGC 4151            &  $ 0.40\pm0.07$  &  $ 4.3  \pm0.6  $  &  $82    \pm9\phantom{2}    $  &  $12    \pm2\phantom{1}    $  &  $240    \pm30\phantom{2}    $   \\
NGC 4258            &  $-0.21\pm0.06\phantom{-}$  &  $ 0.47 \pm0.03 $  &  $ 1.8  \pm0.2  $  &  $ 1.3  \pm0.1  $  &  $ 5.3  \pm0.6  $   \\
NGC 4303            &  $-0.43\pm0.04\phantom{-}$  &  $ 0.43 \pm0.02 $  &  $ 0.92 \pm0.09 $  &  $ 1.2  \pm0.1  $  &  $ 2.7  \pm0.3  $   \\
NGC 4374            &  $-0.86\pm0.04\phantom{-}$  &  $ 1.7  \pm0.1  $  &  $ 0.50 \pm0.15 $  &  $ 4.8  \pm0.3  $  &  $ 1.4  \pm0.4  $   \\
NGC 4388            &  $ 0.44\pm0.09$  &  $ 0.40 \pm0.07 $  &  $ 8.3  \pm1.2  $  &  $ 1.1  \pm0.2  $  &  $24    \pm4\phantom{2}    $   \\
NGC 4395            &  $ 0.25\pm0.09$  &  $ 0.015\pm0.002$  &  $ 0.19 \pm0.03 $  &  $ 0.043\pm0.007$  &  $ 0.53 \pm0.08 $   \\
NGC 4507            &  $ 0.92\pm0.02$  &  $ 3.5  \pm1.7  $  &  $220    \pm20\phantom{2}  $  &  $10    \pm5\phantom{1}    $  &  $630    \pm60\phantom{0}    $   \\
NGC 4552            &  $-0.74\pm0.05\phantom{-}$  &  $ 1.1  \pm0.1  $  &  $ 0.74 \pm0.18 $  &  $ 3.3  \pm0.2  $  &  $ 2.1  \pm0.5  $   \\
NGC 4736            &  $-0.70\pm0.07\phantom{-}$  &  $ 0.11 \pm0.01 $  &  $ 0.083\pm0.024$  &  $ 0.30 \pm0.03 $  &  $ 0.24 \pm0.07 $   \\
MARKARIAN 231       &  $ 0.07\pm0.09$  &  $13    \pm2\phantom{2}    $  &  $100    \pm20\phantom{2}  $  &  $38    \pm5\phantom{2}    $  &  $300    \pm50\phantom{0}    $   \\
NGC 5135            &  $-0.67\pm0.06\phantom{-}$  &  $ 5.6  \pm0.4  $  &  $ 5.2  \pm1.1  $  &  $16    \pm1\phantom{2}    $  &  $15    \pm3\phantom{1}    $   \\
NGC 5194            &  $-0.67\pm0.05\phantom{-}$  &  $ 0.32 \pm0.02 $  &  $ 0.29 \pm0.06 $  &  $ 0.92 \pm0.07 $  &  $ 0.85 \pm0.16 $   \\
CIRCINUS GALAXY     &  $ 0.16\pm0.10$  &  $ 0.11 \pm0.02 $  &  $ 1.0  \pm0.2  $  &  $ 0.30 \pm0.05 $  &  $ 3.0  \pm0.5  $   \\
NGC 5506            &  $ 0.46\pm0.09$  &  $81    \pm16   $  &  $400    \pm50\phantom{2}  $  &  $230    \pm50\phantom{2}  $  &  $1100    \pm200\phantom{0}    $   \\

\enddata
\end{deluxetable*}

\end{document}